\documentclass[12pt]{iopart}
\usepackage{graphicx}
\usepackage{colordvi}
\begin{document}
\title[Effective potential and geodesics in the Wyman geometry]
{Qualitative and quantitative features of orbits of massive 
particles and photons moving in Wyman geometry.}

\author{G Oliveira-Neto$^1$ and G F Sousa$^2$} 

\address{$^1$ Departamento de Matem\'{a}tica e Computa\c{c}\~{a}o,
Faculdade de Tecnologia, \\
Universidade do Estado do Rio de Janeiro,
Rodovia Presidente Dutra Km 298,\\ P\'{o}lo Industrial,
CEP 27537-000, Resende-RJ, Brazil.}
\address{$^2$ Departamento de F\'{\i}sica, Instituto de Ci\^{e}ncias Exatas,\\
Universidade Federal de Juiz de Fora,
CEP 36036-330, Juiz de Fora,\\ Minas Gerais, Brazil.}
\eads{\mailto{gilneto@fat.uerj.br},\mailto{gilberto\_freitas@yahoo.com.br}}

\begin{abstract}
The Wyman's solution depends on two parameters, the mass $M$ and 
the scalar charge $\sigma$. If one fixes $M$ to a positive value, 
say $M_0$, and let $\sigma^2$ take values along the real line it
describes three different types of spacetimes. For $\sigma^2 >0$ 
the spacetimes are naked singularities, for $\sigma^2 = 0$ one has 
the Schwarzschild black hole of mass $M_0$ and finally for 
$-M_0^2 \leq \sigma^2 < 0$ one has wormhole spacetimes. In the 
present work, we shall study qualitative and quantitative features 
of orbits of 
massive particles and photons moving in the naked singularity and 
wormhole spacetimes of the Wyman solution. These orbits are the 
timelike geodesics for massive particles and null geodesics for
photons. Combining the four geodesic equations with an additional
equation derived from the line element, we obtain an effective
potential for the massive particles and a different effective 
potential for the photons. We investigate all possible types of 
orbits, for massive particles and photons, by studying the 
appropriate effective potential. We notice that for certain values
of $\sigma^2 >0$, there is an infinity potential wall that prevents 
both massive particles and photons ever to reach the naked singularity. 
We notice, also, that for certain values of $-M_0^2 \leq \sigma^2 < 0$, 
the potential is finite everywhere, which allows massive particles
and photons moving from one wormhole asymptotically flat region to the 
other. We also compute the radial timelike and null geodesics for
massive particles and photons, respectively, moving in the naked 
singularities and wormholes spacetimes.
\end{abstract}


\submitto{\CQG}

\maketitle

\section{Introduction}
\label{sec:intro}

The {\it weak equivalence principle} of general relativity tell
us that massive particles move along timelike geodesics and
photons move along null geodesics \cite{wheeler}. Despite their
fundamental importance as one of the principles of general 
relativity, the geodesics also help us learning more about
different properties of a given spacetime. A textbook example
comes from the study of geodesics in Schwarzschild geometry 
\cite{wheeler}. Without actually computing the geodesics, just 
observing the effective potential diagram, one can see that 
both massive particles and photons can never leave the event horizon 
once they enter that surface. This is the case because the 
effective potential for both massive particles and photons
diverges to negative infinity as one approaches the singularity
located at the coordinate system origin. Therefore, once the
massive particles and photons enter the event horizon they are
accelerated toward the singularity without any chance to turn
back. Many authors, over the years, have computed the 
effective potential diagram and the geodesics of different
spacetimes in order to learn more about their properties. In
particular, we may mention some important works dealing with 
different black hole spacetimes \cite{blackholegeodesics}.
Two other important gravitational configurations besides black
holes that may form due to the gravitational collapse are
naked singularities and wormholes. Some authors have already
investigated some of their properties by computing 
effective potential diagrams and geodesics for those
space-times \cite{nsgeodesics1,nsgeodesics,wormholegeodesics}.
A well-known spacetime geometry which may describe naked 
singularities as well as wormholes is the Wyman one \cite{wyman}. 
Since, to the best of our knowledge, nobody has ever studied the
Wyman spacetime by means of the computation of its 
effective potential diagrams and geodesics, we decided
to do that in the present work.

The Wyman geometry describes the four dimensional space-time generated 
by a minimally coupled, spherically symmetric, static, massless scalar 
field and has been studied by many authors 
\cite{wyman,bergman,roberts,jetzer,clayton,gil}. From a particular 
case of the general Wyman's solution, M. D. 
Roberts showed how to construct a time dependent solution 
\cite{roberts}. The Roberts' solution has an important physical 
interest because it may represent the gravitational collapse of 
a scalar field. Later, P. Brady and independently Y. Oshiro et 
al. \cite{brady}, \cite{oshiro} showed that the Roberts' solution 
could be derived from the appropriated, time-dependent, 
Einstein-scalar equations by using a continuous self-similarity. They 
also showed that the Roberts' solution exhibits a critical behavior 
qualitatively identical to the one found numerically by M. W. 
Choptuik \cite{choptuik}, studying the same system of equations.
In fact, the above results confirmed early studies of 
D. Christodoulou who pioneered analytical studies of that model 
\cite{chris}.

The Wyman's solution is not usually thought to be of great 
importance for the issue of gravitational collapse because it
is static and the naked singularities derived from it are 
unstable against spherically symmetric linear perturbations of 
the system \cite{jetzer,clayton}. On the other hand, as we saw 
above, from a particular case of the Wyman's solution one may 
derive the Roberts' one which is of great importance for the 
issue of gravitational collapse. Also, it was shown that there 
are nakedly singular solutions to the static, massive scalar 
field equations which are stable against spherically symmetric 
linear perturbations \cite{clayton}. Therefore, we think it is 
of great importance to gather as much information as we can 
about the Wyman's solution for they may be helpful for a better 
understand of the scalar field collapse.

The Wyman's solution depends on two parameters, the mass $M$ and 
the scalar charge $\sigma$. If one fixes $M$ to a positive value, 
say $M_0$, and let $\sigma^2$ take values along the real line it
describes three different types of spacetimes. For $\sigma^2 >0$ 
the spacetimes are naked singularities, for $\sigma^2 = 0$ one has 
the Schwarzschild black hole of mass $M_0$ and finally for 
$-M_0^2 \leq \sigma^2 < 0$ one has wormhole spacetimes. Therefore,
we have an interesting situation where we can study different 
properties of naked singularities and wormholes together through 
the effective potential diagrams and geodesics of massive
particles and photons in the Wyman's solution.

In the next Section, we introduce the Wyman solution and identify
the values of the parameters $M$ and $\eta$ that describes naked
singularities, wormholes and the Schwarzschild black hole. In 
Section \ref{timelike}, we combine the geodesic equations to obtain 
the effective potential equation for the case of massive particles.
We study the effective potential and qualitatively describe the types
of orbits for massive particles moving in the naked singularities and
wormholes spacetimes. In this section, we also compute the radial 
timelike geodesics for particles moving in the naked singularities and
wormholes spacetimes. In Section \ref{null}, we combine the geodesic 
equations to obtain the effective potential equation for the case of 
photons. We study the effective potential and qualitatively describe 
the types of orbits for photons moving in the naked singularities and
wormholes spacetimes. In this section, we also compute the radial 
timelike geodesics for photons moving in the naked singularities and
wormholes spacetimes. Finally, in Section \ref{sec:conclusions} we 
summarize the main points and results of our paper.

\section{Naked singularities, wormholes and the Schwarzschild black hole.}
\label{naked}

The Wyman line element and the scalar field expression are given in terms 
of the coordinates ($t$, $r$, $\theta$, $\phi$) by the equation (9) of 
\cite{roberts},

\begin{equation}
\label{1}
ds^2=-\Big(1-\frac{2\eta}{r}\Big)^{\frac{M}{\eta}}dt^2+
\Big(1-\frac{2\eta}{r}\Big)^{-\frac{M}{\eta}}dr^2+
\Big(1-\frac{2\eta}{r}\Big)^{1-\frac{M}{\eta}}r^2d\Omega^2,
\end{equation}
where $r$ varies in the range $2\eta < r < \infty$ and $d\Omega^2$ 
is the line element of the two-dimensional sphere with unitary 
radius. The scalar field is,
\begin{equation}
\label{2}
\varphi=\frac{\sigma}{2\eta}\ln\Big(1-\frac{2\eta}{r}\Big).
\end{equation}
where $\sigma$ is the scalar charge given by 
$\sigma^2 = \eta^2 - M^2$ and $\sigma^2 \geq -M^2$. We are
working in the unit system where $G = c = 1$.

In the line element (\ref{1}), the function $R(r)$,

\begin{equation}
\label{3}
R(r)=r\Big(1-\frac{2\eta}{r}\Big)^{\frac{1}{2}(1-\frac{M}{\eta})},
\end{equation}
is the physical radius which gives the circumference and area of 
the two-spheres present in the Wyman geometry. Instead o using the
parameters $M$ and $\sigma$ to identify the different types
of spacetimes described by (\ref{1}), we may also use 
the parameters $M$ and $\eta$, since they are all related by the 
equation just below (\ref{2}). Each different type of spacetime
will have a different behaviour of the function $R(r)$. $R(r)$ has
a single extreme point which is a minimum located at $r = M + \eta$.
Since, the line element (\ref{1}) is valid for $r > 2\eta$, only in
the cases where $M > \eta$ there will be a minimum inside the domain
of $r$. Based on that result we have three different cases for a 
positive $M$. As a matter of simplicity we shall use, also, the 
positive parameter $\lambda \equiv M/\eta$ in order to identify 
the three different cases. 

\emph{Case} $M < \eta$ or $0 < \lambda < 1$.

In this case we have the following important values of $R(r)$ 
(\ref{3}),

\begin{equation}
\label{4} 
\lim_{r \rightarrow 2\eta_{+}}R(r)=0,
\qquad \lim_{r \rightarrow \infty}R(r)=\infty.
\end{equation}

Here, the solution represents space-times with a physical 
naked timelike singularity located at $R = 0$ \cite{bergman}. 
It is easy to see that this singularity is physical
because, the Ricci scalar $\mathbf{R}$ computed from the line 
element (\ref{1}),

\begin{equation}
\label{4,5}
\mathbf{R}=\frac{2(M^2-\eta^2)}{r^4}
\Big(1-\frac{2\eta}{r}\Big)^{\frac{M}{\eta}-2},
\end{equation}
diverges there. It is also easy to see that this singularity
is naked with the aid of the quantity
$Q=g^{\alpha\beta}\frac{\partial R}{\partial
x^\alpha}\frac{\partial R}{\partial x^\beta}$. The roots of $Q$
determine the presence of event horizons in spherical symmetric
spacetimes \cite{brady}. For the Wyman solution $Q$ is given by,

\begin{equation}
\label{5}
Q=\Big(1-\frac{2\eta}{r}\Big)^{-1}\Big(1-\frac{M+\eta}{r}\Big)^2.
\end{equation}
The only root of $Q$ is located at $r = M + \eta$ which confirms
that the singularity located at $R=0$ is naked for all spacetimes
in the present case.

This singularity is sometimes called a `central singularity'
and is similar to that appearing in the `extreme'
Reissner-Nordstr{\o}m black hole and in the negative mass 
Schwarzschild spacetime. From equation (\ref{2}), the scalar 
field vanishes asymptotically as $R \to \infty$ and diverges at 
the singularity.

\emph{Case} $M = \eta$ or $\lambda = 1$.

In this case we have $R(r)=r$ and $\sigma=0$. This last condition 
implies that the scalar field (\ref{2}) vanishes and one gets 
the Schwarzchild solution. Here, the minimum of $R(r)$, located at 
$R=2M$, is outside the domain of $r$ and the line element (\ref{1})
describes only the spacetime exterior to the event horizon.

\emph{Case} $M > \eta$ or $1 < \lambda < \infty$.

In this case we have the following important values of $R(r)$ 
(\ref{3}),

\begin{equation}
\label{6}
\lim_{r \rightarrow 2\eta_{+}}R(r)=\infty, \quad
R_{min}=(M+\eta)\Big(\frac{M-\eta}{M+\eta}\Big)^{\frac{1}{2}(1-\frac{M}{\eta})},
\quad \lim_{r \rightarrow \infty}R(r)=\infty.
\end{equation}

Due to the fact that $\sigma^2 \geq -M^2$ as stated
just before (\ref{2}), we have that in the present
case $\eta \geq 0$. The case $\eta = 0$ is well know 
in the literature as the Yilmaz-Rosen space-time 
\cite{rosen}. In this case $M > \eta$, the physical radius 
$R$ is never zero. If one starts with a large value of 
$R$, for large values of $r$, and starts diminishing $R$, 
reducing the values of $r$, one will reach the minimum 
value of $R$ ($R_{min}$ (\ref{6})) for $r = M + \eta$. 
Then, $R$ starts to increase again when we let $r$ goes to 
$2\eta$ until it diverges when $r = 2\eta$. Therefore, we
may interpret these spacetimes as wormholes connecting two
asymptotically flat regions such that they have a minimum
throat radius given by $R_{min}$ (\ref{6}) \cite{gil}.
The spatial infinity ($R \to \infty$) of each asymptotically 
flat region is obtained, respectively, by the limits: 
$r \to \infty$ and $r \to 2\eta_+$.
An important property of this space-time is that the scalar 
field (\ref{2}) is imaginary. The imaginary scalar 
field also known as ghost Klein-Gordon field 
\cite{hayward} is an example of the type of matter 
called {\it exotic} by some authors \cite{thorne}. It 
violates most of the energy conditions and is repulsive. 
This property helps explaining the reason why the 
collapsing scalar field never reaches $R=0$.

As mentioned above, in the rest of the paper we shall restrict
our attention to the spacetimes representing naked singularities
and wormholes.

\section{Timelike Geodesics}
\label{timelike}

\subsection{Effective Potential}
\label{subsectionpotential}

We have four geodesic equations, one for each coordinate 
\cite{wheeler},

\begin{equation}
\label{7}
\frac{d^2x^\alpha}{d\tau^2}+\Gamma^{\alpha}_{\phantom{i}\beta\gamma}
\frac{dx^\beta}{d\tau}\frac{dx^\gamma}{d\tau}=0\, ,
\end{equation}
where $\alpha = 0, 1, 2, 3$, and $x^\alpha$ represents, respectively, 
each of the coordinates ($t$, $r$, $\theta$, $\phi$). $\tau$ is the proper 
time of the massive particle which trajectory is described by 
(\ref{7}).

The geodesic equation in the coordinate $\theta$ tell us that, like in the 
Schwarzschild case, the geodesics are independent of $\theta$, therefore we 
choose the equatorial plane to describe the particle motion ($\theta = \pi/2$). 
In the equatorial plane, the geodesic equation for $\phi$ may be once integrate 
to give,

\begin{equation}
\label{8}
r^2 \Big(1-\frac{2\eta}{r}\Big)^{1-\frac{M}{\eta}}
\dot{\phi}\, =\, R^2\dot{\phi}\, =\, L\, ,
\end{equation}
where the dot means derivative with respect to $\tau$ and $L$ is the integration
constant that may be interpreted as the particle angular momentum per unit rest 
mass. This result means that $\phi$ is {\it cyclic} and its conjugated momentum 
($p_\phi = \dot{\phi}$) is a conserved quantity. Also, $\dot{\phi}$ may be 
written as a function of $r$. Likewise, in the equatorial plane, the geodesic 
equation for $t$ may be once integrate to give,

\begin{equation}
\label{9}
\Big(1-\frac{2\eta}{r}\Big)^{\frac{M}{\eta}}\dot{t}\, =\, E\, ,
\end{equation}
where $E$ is the integration constant that may be interpreted as the particle 
energy per unit rest mass. This result means that $t$ is {\it cyclic} and its 
conjugated momentum ($p_t = \dot{t}$) is a conserved quantity. Also, $\dot{t}$ 
may be written as a function of $r$. Instead of using the fourth geodesic 
equation for the $r$ coordinate, we use the equation derived directly from the 
line element (\ref{1}), $ds^2/d\tau^2 = -1$ \cite{dinverno}. There, we introduce 
the expressions of $\dot{\phi}$ (\ref{8}) and $\dot{t}$ (\ref{9}), in 
order to obtain the following equation which depends only on $r$,

\begin{equation}
\label{10}
\left({dr\over d\tau}\right)^2\, +\, V(r)^2\, =\, E^2\, .
\end{equation}
Where
\begin{equation}
\label{11}
V(r)^2=\Big(1-\frac{2\eta}{r}\Big)^{\frac{M}{\eta}}\Bigg[1+\frac{L^2}{r^2}
\Big(1-\frac{2\eta}{r}\Big)^{\frac{M}{\eta}-1}\Bigg]
\end{equation}
and $V(r)$ is the effective potential for the motion of massive particles 
in the Wyman geometry. The geodesic equation for the $r$ coordinate plays 
the role of a control equation, where we substitute the solutions found with 
the other four equations in order to verify their correctness.

In order to have a qualitative idea on the orbits of massive particles
moving under the action of $V(r)^2$ it is important to draw it as a function
of $r$. We may do it by, initially, computing the extremes of $V(r)^2$. First,
we calculate the first derivative of $V(r)^2$ (\ref{11}) and find the
roots of the resulting equation. That equation may be simplified by the 
introduction of the auxiliary quantities: $x = (1-2\eta/r)$, $0 < x < 1$; 
$A = \eta^2 /L^2$, $0 < A < \infty$; $B = (\lambda - 1/2)/(\lambda + 1/2)$,
$-1 < B < 1$; $C = 2\lambda /(\lambda + 1/2)$, $0 < C < 2$. Where $\lambda$
was defined before and the domains of $A$, $B$ and $C$ where determined by
the fact that $\lambda$ is positive. The equation $dV(r)/dr = 0$, in terms
of $x$, $\lambda$, $A$, $B$ and $C$ is given by,

\begin{equation}
\label{12}
ACx^{1-\lambda}\, +\, B\, -\, Cx\, +\, x^2\, =\, 0.
\end{equation}
In order to identify the presence and nature of the roots of (\ref{12}),
let us define the following two auxiliary functions,

\begin{equation}
\label{13}
p(x)\, =\, -\, x^2\, +\, Cx\, -\, B\,, \qquad
h(x)\, =\, AC x^{1-\lambda}\, .
\end{equation}
Now, the values of $x$ where the two curves $p(x)$ and $h(x)$ meet will
be the roots of (\ref{12}). $p(x)$ is a set of parabolas which vertices
are all located above the x-axis and with two roots. The larger one is located
at $x=1$ and the smaller one is located at $x=(\lambda - 1/2)/(\lambda + 1/2)$.
For $\lambda > 1/2$, the smaller root is positive; for $\lambda = 1/2$, it is
zero and for $\lambda < 1/2$, it is negative. The precise nature of $h(x)$ will 
depend on the value of $\lambda$, present in the exponent of $x$, but whatever 
$\lambda$ one chooses, all values of $h(x)$ will be located above the x-axis.
For $\lambda > 1$, $h(x)$ diverges to $+\infty$ when $x \to 0$ and goes to zero
when $x \to +\infty$. For $\lambda < 1$, $h(x)$ goes to zero when $x \to 0$ and
diverges to $+\infty$ when $x \to +\infty$.

An important root of (\ref{12}) is defined by the value of $x$, say $x_0$,
where the two curves $p(x)$ and $h(x)$ (\ref{13}) just touch each other.
$x_0$ is an inflection point of $V(r)^2$ (\ref{11}), because there the second
derivative of $V(r)^2$ is also zero. Associated to $x_0$ there is a value of the
angular momentum $L$, say $L_0$, which originates an unstable particle orbit.
Due to the fact that $L$ is present, only, in the denominator of $A$ in the
expression of $h(x)$ (\ref{13}), if one increases $L$, $h(x)$ will assume
smaller values for the same values of $x$. $p(x)$ will not be altered because
$L$ is not present in its expression. Therefore, $L_0$ is the value of $L$ for
which $h(x)$ just touches $p(x)$, if one takes values of $L$ greater than $L_0$,
$h(x)$ will start intercepting $p(x)$ in two or more points. They will be
extremes of $V(r)^2$, maximum, minimum or inflection points. It is possible to
compute the value of $L_0$ in terms of $x_0$ and the parameters $\lambda$ and
$\eta$. In order to do that, we consider, initially, the fact that the first 
derivatives of $h(x)$ and $p(x)$ (\ref{13}), in $x=x_0$, are equal and 
express $ACx_0^{-\lambda}$ in terms of other quantities. Then, we use the 
fact that $x_0$ is a root of (\ref{12}) and re-write that equation for
$x=x_0$ and substitute there the value of $ACx_0^{-\lambda}$ just obtained. It
gives rise to the following second degree polynomial equation in $x_0$,

\begin{equation}
\label{14}
x_{0}^{2}-\frac{C\lambda}{\lambda+1}x_{0}+\frac{\lambda-1}{\lambda+1}B=0,
\end{equation}
where $\lambda \neq 1$. It has the following roots,

\begin{equation}
\label{15}
x^{+}_0=\frac{2\lambda^2 +
\sqrt{5\lambda^2-1}}{2(\lambda+1/2)(\lambda+1)}, \qquad
x^{-}_0=\frac{2\lambda^2 -
\sqrt{5\lambda^2-1}}{2(\lambda+1/2)(\lambda+1)},
\end{equation}
where $x^{+}_0 \geq x^{-}_0$.
Due to the fact that we have two distinct values of $x_0$, $x^{+}_0$ and $x^{-}_0$ 
given by (\ref{15}), we shall have, also, two distinct values of $L_0$, say
$L_{0+}$ and $L_{0-}$. In order to obtain them, we introduce $x^{+}_0$ and 
$x^{-}_0$, separately, in the equation that equates the first derivatives of
$h(x)$ and $p(x)$ and express $L_0^2$ in term of other quantities. It results
in,

\begin{equation}
\label{16}
L^2_{0+}=\frac{\eta^{2}\lambda(1-\lambda)}{(x^{+}_{0})^{\lambda}
[\lambda- x^{+}_{0}(\lambda+1/2)]},\qquad
L^2_{0-}=\frac{\eta^{2}\lambda(1-\lambda)}{(x^{-}_{0})^{\lambda}
[\lambda-x^{-}_{0}(\lambda+1/2)]} ,
\end{equation}
where $L^2_{0-} \geq L^2_{0+}$.

By definition $\lambda$ has the domain $[0, \infty )$, but if one
introduces values of $\lambda$, over all this domain, in the expressions of
$x^{+}_{0}$, $x^{-}_{0}$ (\ref{15}) and $L^2_{0+}$, $L^2_{0-}$ (\ref{16}),
one will find some unphysical results. This means that one will have to impose
some restrictions on the domain of $\lambda$. These restrictions will have to
take in account the following two conditions: (i) $x^{+}_{0}$ and $x^{-}_{0}$ 
varies in the range $[0, 1]$, due to the definition of $x$; (ii) $L^2_{0+}$ and 
$L^2_{0-}$ must be positive. These conditions lead to the following distinct 
domains of $\lambda$, depending on the extreme $x_0^+$ or $x_0^-$ one is using,

\begin{equation}
\label{17}
x^{-}_0:\qquad \frac{1}{\sqrt{5}}\, \leq\, \lambda\, <\, \frac{1}{2},
\end{equation}
\begin{equation}
\label{18}
x^{+}_0:\qquad \lambda\, \geq\, \frac{1}{\sqrt{5}}.
\end{equation}

Therefore, this result tell us that $V(r)^2$ (\ref{11}) behaves differently,
depending on the value of $\lambda$. We have the following three different 
regions:

(i) $\lambda \geq 1/2$.

There will be just one inflection point located at $x^{+}_0$ for $L^2_{0+}$. 
If one chooses values of $L^2 > L^2_{0+}$, one will find other extremes of 
$V(r)^2$ (\ref{11}), which will be maximum or minimum points.

(ii) $1/\sqrt{5} \leq \lambda < 1/2$.

There may be two different inflection points. The first located at $x^{+}_0$ 
for $L^2 = L^2_{0+}$. If one chooses values of $L^2 > L^2_{0+}$, one will find 
other extremes of $V(r)^2$ (\ref{11}), which will be maximum or minimum 
points. When one reaches $L^2 = L^2_{0-}$, one finds the other possible 
inflection point located at $x^{-}_0$. If one chooses values of $L^2 > L^2_{0-}$, 
one will find just one extreme of $V(r)^2$ (\ref{11}), which will be a 
minimum point. Even for the case $L^2 < L^2_{0+}$, there will be a minimum point. 
In fact, this minimum point will always be present for any value of $L$. Its 
presence can be understood because, here, $p(x)$ has a negative root and $h(x)$ 
is crescent and starts from $x=0$. Therefore, these two curves will always 
intercept each other.

(iii) $\lambda < 1/\sqrt{5}$.

There will be no inflection points but there will be always a minimum point. The 
presence of this minimum point can be understood in the same way as the one in 
the previous case.

As we saw in the previous section, Sec. \ref{naked}, naked singularities and 
wormholes are characterized by certain subdomains of $\lambda$. Therefore,
based on the above result, we may draw the effective potentials for naked 
singularities and wormholes. From these effective potentials, we shall be able
to describe qualitatively the different orbits of massive particles.

\subsection{Effective Potential for Naked Singularities}
\label{subsectioneffectivenakedparticles}

As we saw, in Sec. \ref{naked}, the naked singularities are obtained for 
$0 < \lambda < 1$. We may still divide the naked singularities in two
classes, due to the behavior of $V(r)^2$ (\ref{11}) when $r \to 2\eta_+$.
For $0 < \lambda < 1/2$, $V(r)^2$ (\ref{11}) assumes the following values
when $r \to 2\eta_+$ and $r \to +\infty$,

\begin{equation}
\label{18,1}
\lim_{r\rightarrow 2\eta_+}V(r)^2=\infty ,\qquad 
\lim_{r\rightarrow +\infty}V(r)^2=1.
\end{equation}
Due to the fact that these limits are consistent with an asymptotically flat
naked singularity located at $r = 2\eta$, we call this class of ordinary
naked singularities. On the other hand, for $1/2 \leq \lambda < 1$ the limit
of $V(r)^2$ (\ref{11}) when $r \to 2\eta_+$ is zero. Due to fact that
this result is not consistent with a naked singularity located at $r \to 2\eta$, 
we call this class of {\it anomalous naked singularities}. Then, in what follows 
we shall restrict our attention to the class of ordinary naked singularities 
with $0 < \lambda < 1/2$. It is important to mention that observing the scalar 
field expression (\ref{2}), which in this case is real, one can see that it
diverges to $-\infty$ as $r \to 2\eta_+$.

Taking in account the results of Subsec. \ref{subsectionpotential} (ii) and (iii)
the effective potential $V(r)^2$ (\ref{11}) may have several different shapes
depending on the value of $L^2$. Here, the inflection points will be located
at $x_0^-$ and $x_0^+$ (\ref{15}) and the relevant angular momenta will be 
$L^2_{0-}$ and $L^2_{0+}$ (\ref{16}).

\begin{itemize}
\item For $L^2 < L^2_{0+}$, $V(r)^2$ (\ref{11}) has one minimum point. In terms
of $x$ it is located in the range ($0$,$x_0^-$) or in terms of $r$ it is located
in the range ($2\eta$ , $r_0^- \equiv 2\eta /(1-x_0^- )$). For $E^2 < 1$, the massive particles
orbit around the naked singularity. If the massive particle is located exactly at
the minimum point the orbit is circular and stable. For $E^2 > 1$, the massive 
particles come in from infinity reach the infinity potential wall near the naked 
singularity and return to infinity without ever reach the naked singularity. 

\item For $L^2 = L^2_{0+}$, $V(r)^2$ (\ref{11}) has two extreme points, a 
minimum located in the range ($2\eta$ , $r_0^-$) and an inflection
point located at $r = r_0^+ \equiv 2\eta /(1-x_0^+ )$. In this point the massive 
particles have unstable circular orbits. The other possible trajectories for
the massive particles are exactly as in the previous case.

\item For $L^2_{0+} < L^2 < L^2_{0-}$, $V(r)^2$ (\ref{11}) has three extreme 
points, a minimum located in the range ($2\eta$ , $r_0^-$), a maximum 
located in the range ($r_0^-$ , $r_0^+$) and another maximum located in the range
($r_0^+$ , $\infty$). If the massive particles are exactly in the maximum points 
they have unstable circular orbits. The other possible trajectories for the massive 
particles are exactly as in the first case.

\item For $L^2 = L^2_{0-}$, $V(r)^2$ (\ref{11}) has two extreme points, a 
minimum located in the range ($r_0^+$ , $\infty$) and an inflection point located at 
$r = r_0^-$. In this point the massive particles have unstable 
circular orbits. The other possible trajectories for the massive particles are 
exactly as in the first case.

\item For $L^2 > L^2_{0-}$, $V(r)^2$ (\ref{11}) has one minimum point located 
in the range ($r_0^+$ , $\infty$). The possible trajectories for the massive 
particles are exactly as in the first case.
\end{itemize}

It is important to mention that, for sufficiently large values of $E$, any massive 
particles will come in from infinity and it will get reflected by the infinity 
potential wall near the naked singularity. Then, it will return to infinity 
without ever reach the naked singularity. This result is similar to the one found 
in \cite{nsgeodesics} for timelike geodesics of the naked singularity present in 
the Reissner-Nordstr\"{o}m spacetime. Let us consider, as an example, the case 
where $M=10$ and $\lambda=0.45$. Therefore, we have $L^2_{0+}=954.9308516$ and 
$x^{+}_0=0.1875874406$ which in terms of $r$ is given by $r^{+}_0=54.70674220$. 
We also have, $L^2_{0-}=960.0309665$ and $x^-_0=0.1064234487$ which in terms of 
$r$ is given by $r^{-}_0=49.73770224$. In the Figure 1, we plot five different 
effective potential diagrams from $V(r)^2$ (\ref{11}), one for each of the five 
cases discussed above. They have the five different values of $L^2$: $951$, 
$954.9308516$, $957$, $960.0309665$ and $962$. The naked singularity is
located at $r=44.44444444$.

\begin{figure}[!h]
\label{fig1}
\begin{center}
\includegraphics[width=7cm,height=11cm,angle=-90]{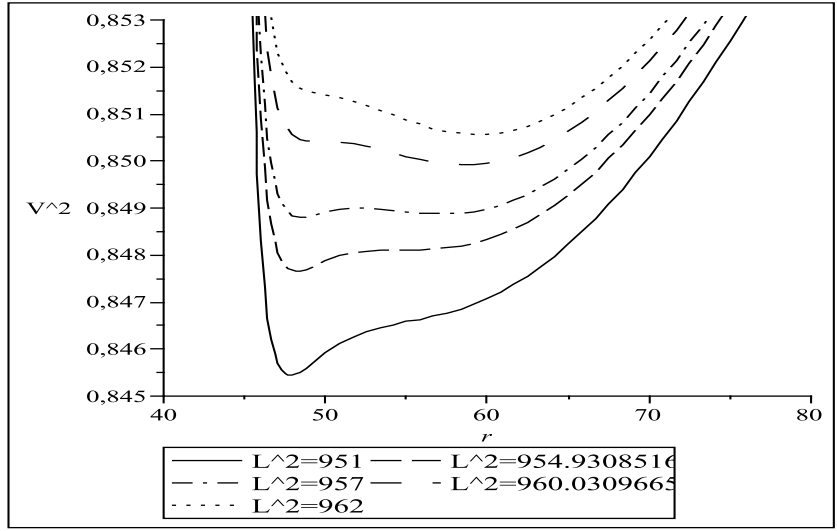}
\caption{Five different effective potential diagrams from $V(r)^2$ 
(\ref{11}), for massive particles moving  in a naked singularity with 
$M=10$ and $\lambda=0.45$. They have the five different values of $L^2$: 
$951$, $954.9308516$, $957$, $960.0309665$ and $962$. The naked 
singularity is located at $r=44.44444444$.}
\end{center}
\end{figure}

\subsection{Effective Potential for Wormholes}
\label{subsectioneffectivewormholeparticles}

As we saw, in Sec. \ref{naked}, the wormholes are obtained for $1 < \lambda < 
\infty$. In the spatial infinity of each asymptotically flat region the 
effective potential $V(r)^2$ (\ref{11}) assumes the following values,

\begin{equation}
\label{19}
\lim_{r\rightarrow 2\eta_+}V(r)^2=0,\qquad \lim_{r\rightarrow \infty}V(r)^2=1.
\end{equation}
Although the above limits give consistent values for the effective potential
at the two spatial infinities, the first limit is not consistent with the
value of the Ricci scalar evaluated at $r=2\eta$, for wormholes with 
$1 < \lambda \leq 2$. For this class of wormholes the Ricci scalar $\mathbf{R}$
(\ref{4,5}), diverges to $\infty$ when we take the limit $r \to 2\eta_+$ or
gives a positive constant when $\lambda = 2$. These values are not consistent 
with the first limit of $V(r)^2$ in (\ref{19}) and the idea of an 
asymptotically flat spatial region. Therefore we call this class of wormholes 
of {\it anomalous wormholes}. For the rest of the wormholes where $2 < \lambda < 
\infty$, the limit of $\mathbf{R}$ (\ref{4,5}) when $r \to 2\eta_+$ is zero.
This value is consistent with the first limit of $V(r)^2$ in (\ref{19})
and the idea of an asymptotically flat spatial region. Then, in what follows 
we shall restrict our attention to that class of ordinary wormholes with $2 < 
\lambda < \infty$. It is important to mention that observing the scalar field
expression (\ref{2}), which in this case is imaginary, one can see that it
diverges to $-\infty$ as $r \to 2\eta_+$. Since there is no physical singularity
at $r \to 2\eta_+$ for the ordinary class or wormholes, this result indicates
that the coordinates we are using might not be suitable to describe this 
asymptotically flat region.

Taking in account the results of Subsec. \ref{subsectionpotential} (i) the 
effective potential $V(r)^2$ (\ref{11}) may have three different shapes
depending on the value of $L^2$. Here, the inflection point will be located
at $x_0^+$ (\ref{15}) and the relevant angular momentum will be $L^2_{0+}$
(\ref{16}).

\begin{itemize}
\item For $L^2 < L^2_{0+}$, $V(r)^2$ (\ref{11}) has no extreme points. In this 
case, there is no stable orbits. For sufficiently high energies the massive 
particles may travel from one asymptotically flat region to the other. In
fact, this type of orbit is also present in the other two cases considered
below.

\item For $L^2 = L^2_{0+}$, $V(r)^2$ (\ref{11}) has one inflection point, located 
at $x_0^+$ (\ref{15}) or in terms of $r$, $r_0^+ \equiv 2\eta /(1-x_0^+)$. In 
this point the massive particles have unstable circular orbits.

\item For $L^2 > L^2_{0+}$, $V(r)^2$ (\ref{11}) has two extreme points, a
maximum located at $2\eta < r < r_0^+$ and a minimum located at
$r_0^+ < r$. There are closed and open orbits depending on the values of the
total energy and angular momentum of the massive particles.
\end{itemize}

Let us consider, as an example, the case where $M=1$ and $\lambda=\sqrt{1000}$. 
Therefore, we have $L^2_{0+}=12.41266$ and $x^{+}_{0}=0.98799$ which in terms of
$r$ is given by $r^{+}_0=5.26747$. In the Figure 2, we plot three different 
effective potential diagrams from $V(r)^2$ (\ref{11}), one for each of the three 
cases discussed above. They have the three different values of $L^2$: $10$, 
$12.41266$ and $14.4$. The spatial infinities of each asymptotically flat 
region are located at $r =0.06325$ and $r \to \infty$. 

\begin{figure}[!h]
\label{fig2}
\begin{center}
\includegraphics[width=7cm,height=11cm,angle=-90]{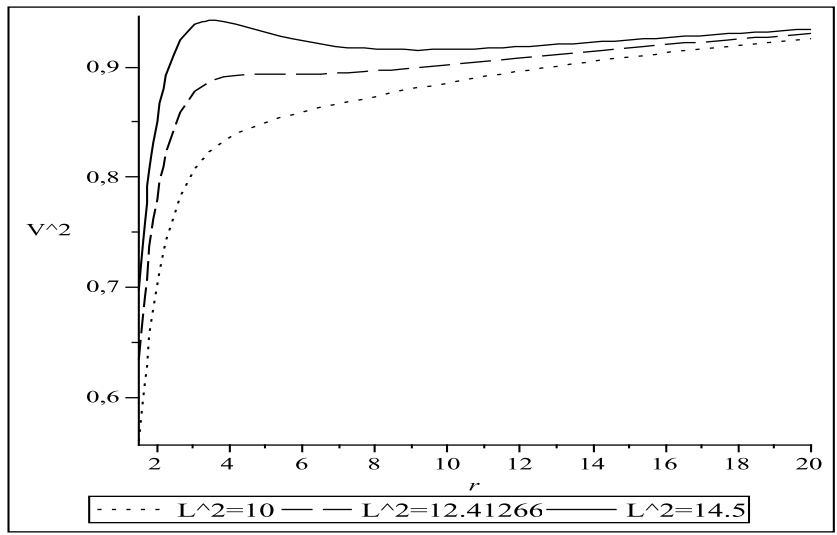}
\caption{Three different effective potential diagrams from $V(r)^2$ (\ref{11}), 
for massive particles moving in a wormhole with $M=1$ and $\lambda=\sqrt{1000}$. 
They have the three different values of $L^2$: $10$, $12.41266$ and $14.4$. 
The spatial infinities of each asymptotically flat region are located at 
$r =0.06325$ and $r \to \infty$.}
\end{center}
\end{figure}

It is important to mention, at this point, that the same study of the qualitative
features of the orbits of massive particles in Wyman geometry in terms of the
proper time $\tau$, as we just did, can also be done in terms of the coordinate
time $t$. In order to do it, we re-write equation (\ref{10}) in the following way,

\begin{equation}
\label{19.05}
\left(E{dr^{*}\over dt}\right)^2\, +\, V(r)^2\, =\, E^2\, ,
\end{equation}
where we used the relationship, $dt/d\tau = E/(1-2\eta/r)^\lambda$, and we introduced
the generalization of the Wheeler's {\it tortoise coordinate} $r^{*}$ \cite{wheeler}, 
given by,

\begin{equation}
\label{19.06}
r^{*} = \int {dr\over \left(1 - {2\eta \over r}\right)^\lambda}\, .
\end{equation}
Since, $V(r)$ and $E$ in (\ref{19.05}) are the same as the ones in (\ref{10}), 
the turning points of $V(r)$, for a given value or $E$, are the same whether we use $t$
or $\tau$. On the other hand, the geodesics may have different properties depending on
the time variable we are using. We shall see some similarities and differences in the 
next section. 

\subsection{Radial timelike geodesics}
\label{radialtimelikegeodesics}

\subsubsection{Description in terms of $\tau$}

Unfortunately, there is not an analytic expression for the general 
timelike geodesics given by the solutions of Eqs. (\ref{8}-\ref{10}),
even the numerical solutions are very complicated. On the other hand,
one may restrict his attention to the case of radial timelike geodesics,
where the massive particle moves only along the radial and time directions.
It means that $\dot{\theta}=\dot{\phi}=L=0$ and (\ref{10}) reduces to,

\begin{equation}
\label{19.1}
\left({dr\over d\tau}\right)^2\, =\, E^2\ -\, 
\Big(1-\frac{2\eta}{r}\Big)^{\frac{M}{\eta}}\, .
\end{equation}

Although (\ref{19.1}) is much simpler than (\ref{10}), one still
cannot integrate it to find an analytic expression of $\tau$ as a function of
$r$. The best one can do is finding a numerical solution for the integral
and plot that solution in a graph of $\tau$ versus $r$. We did that and we
shall present the results below, for the naked singularities and wormholes.

\paragraph{Naked singularities}

In this case, we integrated (\ref{19.1}) for many different values of
$E$, $M$ and $\eta$. We chose values of $M$ and $\eta$ compatibles with
naked singularities. We found that, for $E \geq 1$, the geodesics are all
well behaved when $r \to 2\eta$. In fact, $\tau$ goes to zero in that limit.
The geodesics are such that when $r$ is large, $r$ tends to a linear 
function of the proper time $\tau$. For $E < 1$, the geodesics are also 
well behaved when $r \to 2\eta$ and $\tau$ also goes to zero in that limit.
On the other hand, they do not extend to large values of $r$. It is clear
from (\ref{19.1}), that the massive particle is subjected to a potential
of the form: $(1-2\eta /r)^{{M/ \eta}}$. This potential, increases from zero
at $r=2\eta$ and tends to one when $r\to\infty$. Therefore, massive particles
with a total energy $E < 1$, get reflected by the potential. The value of $r$,
where the particle gets reflected by the potential is obtained by solving the
equation: $E^2=(1-2\eta /r)^{{M/ \eta}}$. Then, based on the above results,
we may conclude that it always takes a finite proper time interval to travel
from any finite value of $r$ to the singularity located at $r=2\eta$. We notice,
also, that for radial timelike geodesics there is not an infinity potential wall 
near the naked singularity and the massive particles can reach it. In the 
Figure 3, we show as an example a geodesic for the case: $E=1$, $M=1$ and
$\eta =3$. The naked singularity is located at $r=6$ and the geodesics extends
to large values of $r$. In the Figure 4, we show as an example a geodesic for 
the case: $E=0.5$, $M=1$ and $\eta =3$. The naked singularity is located at 
$r=6$ and the geodesics gets reflected at $r=6.095238095$. One cannot see the 
geodesic returning to $r=6$, from this figure, because we have considered only 
the outgoing geodesic.

\begin{figure}[!h]
\label{fig3}
\begin{center}
\includegraphics[width=5cm,height=9cm,angle=-90]{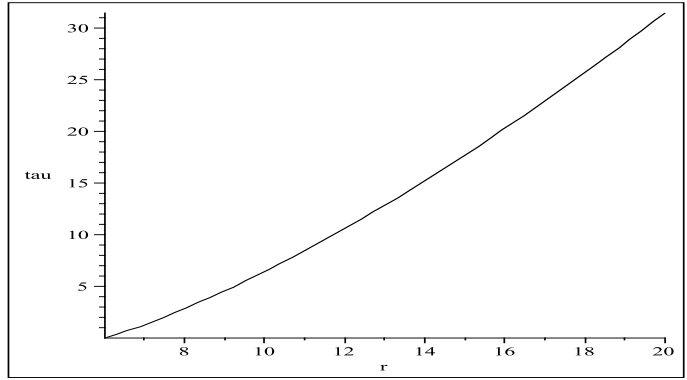}
\caption{Radial geodesic, described in terms of $\tau$, for a massive particle 
moving in a naked singularity with $E=1$, $M=1$ and $\eta =3$. The naked 
singularity is located at $r=6$.}
\end{center}
\end{figure}

\begin{figure}[!h]
\label{fig4}
\begin{center}
\includegraphics[width=5cm,height=9cm,angle=-90]{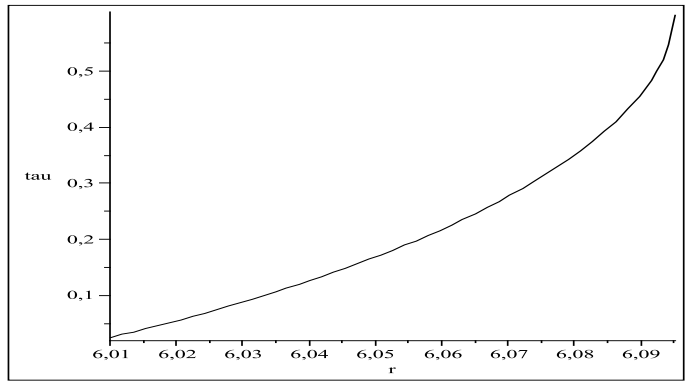}
\caption{Radial geodesic, described in terms of $\tau$, for a massive particle 
moving in a naked singularity with $E=0.5$, $M=1$ and $\eta =3$. The naked 
singularity is located at $r=6$.}
\end{center}
\end{figure}

\paragraph{Wormholes}

In this case, we integrated (\ref{19.1}) for many different values of
$E$, $M$ and $\eta$. We chose values of $M$ and $\eta$ compatibles with
wormholes. We found that, for $E \geq 1$, the geodesics are all
well behaved when $r \to 2\eta$. In fact, $\tau$ goes to zero in that limit.
The geodesics are such that when $r$ is large, $r$ tends to a linear 
function of the proper time $\tau$. For $E < 1$, the geodesics are also 
well behaved when $r \to 2\eta$ and $\tau$ also goes to zero in that limit.
On the other hand, they do not extend to large values of $r$ because they 
get reflected by the potential, as in the naked singularity case. The value 
of $r$, where the particle gets reflected by the potential is obtained in 
the same way we proceeded in the naked singularity case. Here, as in the 
naked singularity case, for all cases studied it always takes a finite 
proper time interval to travel from any finite value of $r$ to the spatial 
infinity located at $r=2\eta$. In the Figure 5, we show as an example a 
geodesic for the case: $E=1$, $M=3$ and $\eta = 1$. The spatial infinities of 
each asymptotically flat regions are located at $r=2$ and $r \to \infty$ and 
the geodesics extends to large values of $r$. 

\begin{figure}[!h]
\label{fig5}
\begin{center}
\includegraphics[width=5cm,height=9cm,angle=-90]{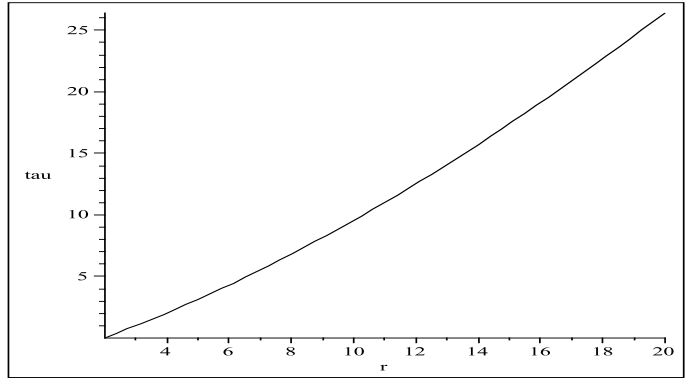}
\caption{Radial geodesic, described in terms of $\tau$, for a massive 
particle moving in a wormhole with $E=1$, $M=3$ and $\eta =1$. The spatial 
infinities of each asymptotically flat regions are located at $r=2$ and 
$r \to \infty$.}
\end{center}
\end{figure}

\subsubsection{Description in terms of $t$}

Here, as in the case where the geodesics are described in terms of $\tau$,
the treatment of the general situation is very complicated. Therefore, we 
shall, also, restrict our attention to the particular situation of radial 
timelike geodesics. From Eqs. (\ref{11}), (\ref{19.05}) and (\ref{19.06}),
appropriately written for the radial case, we may obtain the expression
which, after integration, gives $r$ as a function of $t$.

\begin{equation}
\label{19.5}
t = \int {E\over \sqrt{E^2 - (1 - 2\eta/r)^\lambda }} 
{dr\over (1 - 2\eta/r)^\lambda}
\end{equation}
Even with this simplification, as in the case with $\tau$, we cannot 
integrate (\ref{19.5}) to find an analytic expression of $t$ as a 
function of $r$. The best one can do is finding a numerical solution for 
the integral. We did that for both cases of the naked singularities and 
wormholes. 

\paragraph{Naked singularities}

In this case, we integrated (\ref{19.5}) for many different values of
$E$, $M$ and $\eta$. We chose values of $M$ and $\eta$ compatibles with
naked singularities. Qualitatively, the geodesics behave very much like the
description in terms of $\tau$. For $E \geq 1$ and $E < 1$, $t$ goes to zero 
when $r \to 2\eta$. For $E \geq 1$, when $r$ is large, $r$ tends to a linear 
function of $t$. 
For $E < 1$, the geodesics do not extend to large values of $r$. As we have
mentioned above, since the potential is the same as in the description in 
terms of $\tau$, the turning points are also the same. Nevertheless, there
is a quantitative difference between the two descriptions. We observe that,
although, in both descriptions $r$ is an increasing function of the time 
variable, it grows slower with $t$ than with $\tau$. The specific difference
in the rate of growth between the two descriptions depends on the values of 
the parameters $M$ and $\eta$. Here, we may, also, conclude that it always 
takes a finite time coordinate interval to travel from any finite value of 
$r$ to the singularity located at $r=2\eta$ and there is not an infinity 
potential wall near the naked singularity and the massive particles can 
reach it.

\paragraph{Wormholes}

In this case, we integrated (\ref{19.5}) for many different values of
$E$, $M$ and $\eta$. We chose values of $M$ and $\eta$ compatibles with
wormholes. In analogy with the case where the geodesics are described by 
$\tau$, for $E \geq 1$, when $r$ is large, $r$ tends to a linear function 
of $t$. On the other hand, for $E < 1$ they do not extend to large values 
of $r$. As we have mentioned above, since the potential is the same as in 
the description in terms of $\tau$, the turning points are also the same. 
We found, here, some differences with respect to the description in terms 
of $\tau$. The most important is that, for any value of $E$, the 
geodesics are not well behaved when $r \to 2\eta_+$. In fact, $t$ goes to 
$-\infty$ in that limit. As another difference, we observe that, although, 
in both descriptions $r$ is an increasing function of the time variable, it 
grows slower with $t$ than with $\tau$. The specific difference in the rate 
of growth between the two descriptions depends on the values of the 
parameters $M$ and $\eta$. Due to the divergence in $t$ when $r \to 2\eta_+$, 
we conclude, here, that it always takes an infinite time coordinate interval 
to travel from any finite value of $r$ to the spatial infinity located at 
$r=2\eta$. In the Figure 6, we show as an example a 
geodesic for the case: $E=1$, $M=3$ and $\eta =1$. The spatial infinities of 
each asymptotically flat regions are located at $r=2$ and $r \to \infty$ and 
the geodesics extends to large values of $r$.

\begin{figure}[!h]
\label{fig6}
\begin{center}
\includegraphics[width=5cm,height=9cm,angle=-90]{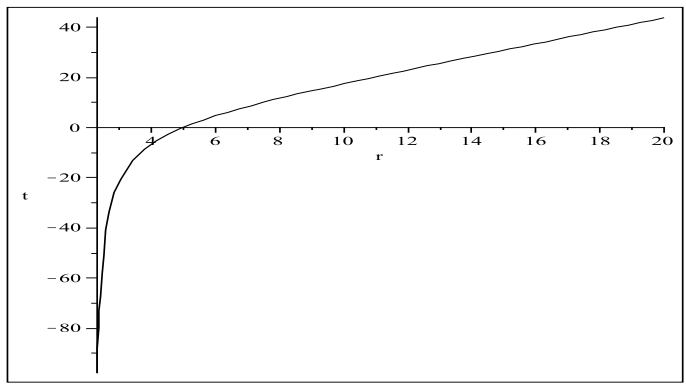}
\caption{Radial geodesic, described in terms of $t$, for a massive particle 
moving in a wormhole with $E=1$, $M=3$ and $\eta =1$. The spatial infinities 
of each asymptotically flat regions are located at $r=2$ and $r \to \infty$.}
\end{center}
\end{figure}

\section{Null Geodesics}
\label{null}

\subsection{Effective Potential}
\label{subsectionpotentialnull}

The null geodesics for the Wyman solution are derived almost in the same way
the timelike geodesics were derived in the previous section. The only difference
is that, here, the null line element contributes a different additional equation
to the four geodesic equations. The new equation reads: $ds^2/d\chi^2 = 0$,
where $\chi$ is the affine parameter used in the present case. Therefore,
proceeding exactly as in the previous section we obtain the following effective
potential equation,

\begin{equation}
\label{20}
\left({dr\over d\chi}\right)^2\, +\, V(r)^2\, =\, {1\over b^2}\, ,
\end{equation}
where
\begin{equation}
\label{21}
V(r)^2=r^{-2}\Big(1-\frac{2\eta}{r}\Big) ^{\frac{2M}{\eta}-1}
\end{equation}
and $b \equiv L/E$ is the photon impact parameter. $V(r)^2$ (\ref{21}) is 
the effective potential for the motion of the photon in Wyman geometry.

$V(r)^2$ (\ref{21}) has only one extreme point at $r = 2M + \eta$. Due to 
the domain of $r$ it can only exists if $2M > \eta$ or $\lambda > 1/2$ and
when it exists it is a maximum point. Therefore, we have three different cases:
(i) $\lambda > 1/2$, the effective potential has a maximum point at 
$r = 2M + \eta$ and goes to zero at $r = 2\eta$;(ii) $\lambda = 1/2$, the 
effective potential has no extreme points and diverges to $\infty$ as 
$r \to 0$; (iii) $\lambda < 1/2$, the effective potential has no extreme 
points and diverges to $\infty$ as $r \to 2\eta$.

As we saw in Sec. \ref{naked}, naked 
singularities and wormholes are characterized by certain subdomains of 
$\lambda$. Therefore, based on the above result, we may draw the effective 
potentials for naked singularities and wormholes. From these effective 
potentials, we shall be able to describe qualitatively the different orbits of 
photons.

\subsection{Effective Potential for Naked Singularities}
\label{subsectioneffectivenakedparticlesnull}

As it was mentioned in Subsec. \ref{subsectioneffectivenakedparticles},
we are only concerned, here, with the class of ordinary naked singularities 
characterized $0 < \lambda < 1/2$. For this class, $V(r)^2$ (\ref{11}) 
assumes the following values when $r \to 2\eta_+$ and $r \to +\infty$,

\begin{equation}
\label{21,5}
\lim_{r\rightarrow 2\eta_+}V(r)^2=\infty ,\qquad 
\lim_{r\rightarrow +\infty}V(r)^2=0.
\end{equation}
$V(r)^2$ (\ref{21}) has no extreme points. In this case, whatever the 
impact parameter $b$ the photons come in from infinity get reflected by the 
infinity potential wall near the naked singularity and return to infinity 
without ever reach the naked singularity.

Let us consider, as an example, the case where $M=0.3$, $\eta = 1$ and 
$\lambda = 0.3$. In the Figure 7, we plot the effective potential diagram 
from $V(r)^2$ (\ref{21}), for these parameter values. The naked
singularity is located at $r=2$.

\begin{figure}[!h]
\label{fig7}
\begin{center}
\includegraphics[width=6cm,height=10cm,angle=-90]{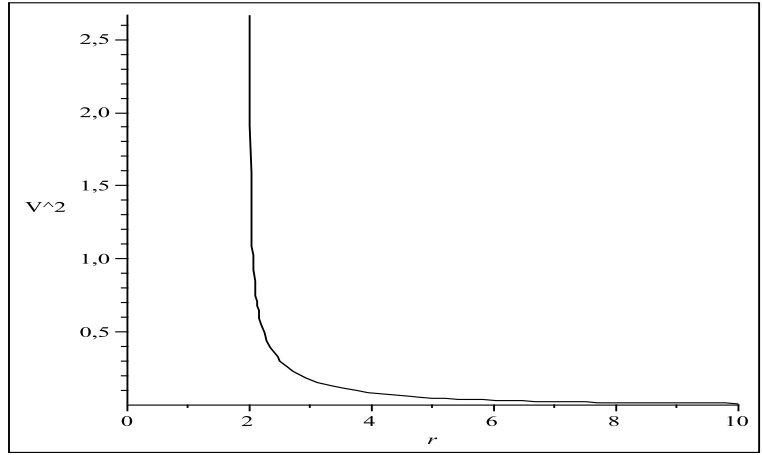}
\caption{Effective potential diagram from $V(r)^2$ (\ref{21}), for 
photons moving in a naked singularity with $M=0.3$, $\eta=1$ and 
$\lambda=0.3$. The naked singularity is located at $r=2$.}
\end{center}
\end{figure}

\subsection{Effective Potential for Wormholes}
\label{subsectioneffectivewormholeparticlesnull}

As it was mentioned in Subsec. \ref{subsectioneffectivewormholeparticles},
we are only concerned, here, with the class of ordinary wormholes characterized
by $2 < \lambda < \infty$. In the spatial infinity of each asymptotically flat 
region the effective potential $V(r)^2$ (\ref{21}) assumes the following 
values,

\begin{equation}
\label{22}
\lim_{r\rightarrow 2\eta_+}V(r)^2=0,\qquad \lim_{r\rightarrow \infty}V(r)^2=0.
\end{equation}
$V(r)^2$ (\ref{21}) has a maximum point located at $r = 2M + \eta$. In 
this point the photons have unstable circular orbits. For sufficiently small 
impact parameter $b$ the photons may travel from one asymptotically flat region 
to the other. Otherwise, they come in from spatial infinity of an asymptotically 
flat region get reflected by the effective potential and return to spatial 
infinity in the same asymptotically flat region.

Let us consider, as an example, the case where $M=30$, $\eta = 1$ and 
$\lambda = 30$. In the Figure 8, we plot the effective potential diagram 
from $V(r)^2$ (\ref{21}), for these parameter values. The spatial infinities of 
each asymptotically flat regions are located at $r=2$ and $r \to \infty$.

\begin{figure}[!h]
\label{fig8}
\begin{center}
\includegraphics[width=6cm,height=10cm,angle=-90]{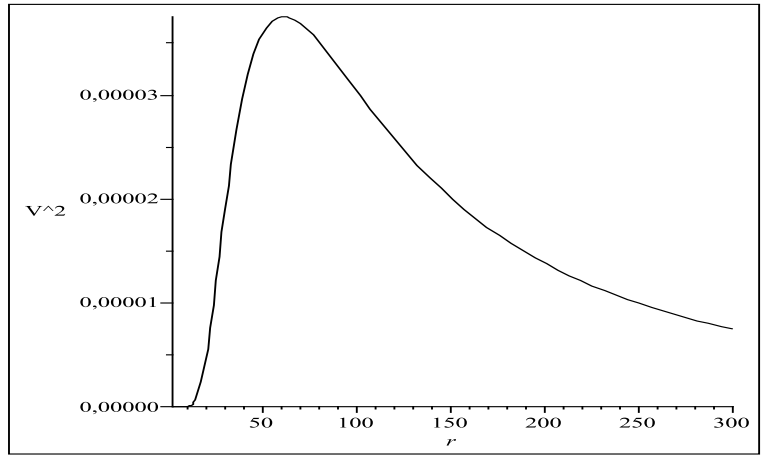}
\caption{Effective potential diagram from $V(r)^2$ (\ref{21}), for photons 
moving in a wormhole with $M=30$, $\eta=1$ and $\lambda=30$. The spatial 
infinities of each asymptotically flat regions are located at $r=2$ and 
$r \to \infty$.}
\end{center}
\end{figure}

\subsection{Radial null geodesics}
\label{radialnullgeodesics}

\subsubsection{Description in terms of $\chi$}

The description of the null geodesics in terms of the affine 
parameter $\chi$ is trivial, when we consider radial geodesics 
($\dot{\theta}=\dot{\phi}=L=0$). That is the case because the
effective potential equation for the radial motion of photons
reduces to $(dr/d\chi)^2=E^2$. It means that the effective 
potential is zero and the photons move freely following straight
lines: $r=\pm E\chi + q$, where $q$ is an integration constant
and the $+$ and $-$ signs mean, respectively, outgoing and ingoing 
geodesics. This result is valid for photons moving in the naked 
singularity as well as in the wormhole geometries.

\subsubsection{Description in terms of $t$}

The description of the null geodesics in terms of the time coordinate 
$t$ may be derived with the aid of (\ref{9}), appropriately
written in terms of $\chi$, and the equation $(dr/d\chi)^2=E^2$.
Then, we have: $(dt/dr)^2=(dt/d\chi)^2/(dr/d\chi)^2=
(1-2\eta/r)^{(-2\lambda)}$. From this equation, we obtain $t$ as a 
function of $r$ after performing the following integral,

\begin{equation}
\label{23}
t = \pm \int \left( 1 - {2\eta\over r}\right)^{-\lambda}dr
\end{equation}

Fortunately, we can analytically integrate (\ref{23}) to find,

\begin{equation}
\label{24}
t=\pm \Bigg[r+2M\ln \Big(\frac{r}{2\eta}\Big)
-\frac{2M(M+\eta)}{r}\,{}_{3}F_{2} \Big(\frac{M}{\eta}+2\,,1\,,1\,
;2\, ,3\, ; \frac{2\eta}{r}\Big) \Bigg] + Q.
\end{equation}
where the $+$ and $-$ signs mean, respectively, outgoing and ingoing 
geodesics. $Q$ is an integration constant and $F$ 
is an hypergeometric function. If we introduce the condition $M=\eta$
in (\ref{24}), we obtain: $t=r+2M\ln(r-2M)+ Q1$,
where $Q1$ is an integration constant. That 
expression is the well-known radial, outgoing null geodesics for the
Schwarzschild geometry parametrized by the time coordinate
\cite{wheeler}.

For large values of $r$ (\ref{24}) indicates that $r$ becomes 
a linear function of $t$, for both naked singularities and wormholes.
On the other hand, a great difference appears between naked 
singularities and wormholes in the limit when $r \to 2\eta_+$. Observing
(\ref{24}), we see that the limit of $t$ when $r \to 2\eta_+$ depends
crucially on the behaviour of ${}_{3}F_{2}$ when $r \to 2\eta_+$. From
the theorem 2.1.2, page 62 in \cite{andrews}, we learn that in
the limit $r \to 2\eta_+$, ${}_{3}F_{2} (M/\eta +2, 1, 1; 2, 3; 2\eta /r)$
converges absolutely if $M/\eta < 1$ and diverges otherwise 
($M/\eta \geq 1$). Therefore, for naked singularities $t$ will converge
for a finite value and for wormholes it will diverge to $-\infty$. It 
is actually possible to compute the limit of 
${}_{3}F_{2} (M/\eta +2, 1, 1; 2, 3; 2\eta /r)$ when $r \to 2\eta_+$. In
order to do that, we must use: equation (2.2.2), page 67 in  
\cite{andrews}, ${}_{2}F_{1} (1\,,1\,;2\,;z)=-\frac{1}{z}\ln|\,1-z\,|$
and $\Gamma(x)\Gamma(1-x)=\pi/\sin(\pi x)$. After an integration we find
the following result,

\begin{equation}
\label{25}
{}_{3}F_{2}(\lambda+2,\,1,\,1;\,2,\,3\,;\,1)=\frac{2\eta}{M+\eta}\Bigg[\,1-\psi
\Bigg(1-\frac{M}{\eta}\Bigg)-\gamma\,\Bigg] \,,
\end{equation}
where $\psi(x) \equiv \Gamma'(x)/\Gamma(x)$ is the digamma function, 
$\Gamma(x)$ is the gamma function and $\gamma = 0.5772156649\ldots$ 
is the Euler constant. Therefore, introducing this result in 
(\ref{24}) we obtain the value of $t$ when $r \to 2\eta_+$,

\begin{equation}
\label{26}
\lim_{r\rightarrow 2\eta_+}t=2(\eta-M)+2M\,\Bigg[\,\psi
\Bigg(1-\frac{M}{\eta}\Bigg)+\gamma\,\Bigg] + Q\, .
\end{equation}

Based on the above results, we may conclude that for the case of naked 
singularities it always takes a finite time coordinate interval to travel 
from any finite value of $r$ to the singularity located at $r=2\eta$.
On the other hand, it always takes an infinite time coordinate interval 
to travel from any finite value of $r$ to the spatial infinity located 
at $r=2\eta$, in the case of wormholes. Both results are qualitatively 
analogous to the corresponding ones found for the radial timelike 
geodesics. Finally, we notice, also, that for null radial geodesics there 
is not an infinity potential wall near the naked singularity and the 
photons can reach it.

\section{Conclusions.}
\label{sec:conclusions}

In the present work, we studied qualitative features of orbits of 
massive particles and photons moving in the naked singularity and 
wormhole spacetimes of the Wyman solution. These orbits are the 
timelike geodesics for massive particles and null geodesics for
photons. Combining the four geodesic equations with an additional
equation derived from the line element, we obtained an effective
potential for the massive particles and a different effective 
potential for the photons. We investigated all possible types of 
orbits, for massive particles and photons, by studying the 
appropriate effective potential. We noticed that for certain values
of $\sigma^2 >0$, there is an infinity potential wall that prevents 
both massive particles and photons ever to reach the naked singularity.
This result is similar to the one found in \cite{nsgeodesics} for 
timelike geodesics of the naked singularity present in the 
Reissner-Nordstr\"{o}m spacetime. 
We noticed, also, that for certain values of $-M_0^2 \leq \sigma^2 < 0$, 
the potential is finite everywhere, which allows massive particles
and photons moving from one wormhole asymptotically flat region to the 
other. We also computed the radial timelike and null geodesics for
massive particles and photons, respectively, moving in the naked 
singularities and wormholes spacetimes.

\ack
G. Oliveira-Neto thanks CNPq and FAPERJ for partial financial support 
and G. F. Sousa thanks CAPES for financial support.

\end{document}